\documentclass[fleqn,twoside]{article}
\usepackage{espcrc2}

\usepackage{graphicx}
\usepackage[figuresright]{rotating}

\newcommand{\AmS}{{\protect\the\textfont2
  A\kern-.1667em\lower.5ex\hbox{M}\kern-.125emS}}

\title{MC generators in CHORUS}

\author{I. Tsukerman \address[ITEP]{ITEP, Moscow, Russia and
CERN, Geneva, Switzerland} for the CHORUS Collaboration}
\begin{document}

\begin{abstract}

This note presents an overview of general-purpose and 
specific Monte-Carlo event generators used in the simulation of
the CERN - CHORUS experiment, aiming to search for
$\nu_{\mu} \rightarrow \nu_{\tau}$ oscillations and charm
particle decays in an emulsion target. 
\end{abstract}
\maketitle
\section{Introduction}
CHORUS is an experiment primarily designed to search for
$\nu_{\mu} \rightarrow \nu_{\tau}$ oscillations 
in the CERN wide-band neutrino beam.  
It also aims to investigate a wide range of 
charm physics.
In four years of exposure, more than 10$^{6}$ neutrino
interactions have been accumulated in the emulsion target,
out of which several thousands lead to the production of a
charmed particle. Using its 
massive calorimeter as a 
target, CHORUS also has a potential to study other
neutrino physics for which large statistics is the crucial point. 

Monte-Carlo (MC) simulation is an essential part of the
experiment. It includes a detailed model of the neutrino beam,
generation of neutrino events, full GEANT-based
simulation of the detector response and the reconstruction. 

This report is organized as follows. 
In Section 2, a brief description of the CHORUS physics
and experimental set-up is given. The MC procedure
is reviewed in Section 3. Section 4 is devoted to
general-purpose event generators and their
verification on the existing experimental data.
In Section 5, more specific generators 
are considered.

\section{CHORUS: physics and set-up}   
The CHORUS detector \cite{Esk97} is a hybrid setup.
A 770-kg nuclear emulsion is used as the primary target
for neutrino interactions, allowing three-dimensional 
reconstruction of short-lived particles like the $\tau$ lepton
and any charmed hadron. 
The emulsion target consists of four stacks.
Each stack is followed by three interface
emulsion sheets 
and by a set of scintillating fibre tracker planes.
They provide accurate predictions of 
particle trajectories into the  emulsion stack for the location of 
vertex positions. The emulsion scanning is  performed
by fully automatic microscopes equipped with CCD cameras and a read-out system. 
The electronic detectors downstream of the emulsion target include a hadron spectrometer 
which measures the bending of charged particles in an air-core magnet, a 100t calorimeter 
where the energy and direction of showers are measured and a muon spectrometer
which determines charge and momentum of muons. 
In addition, the calorimeter was used as an active (lead) target
for neutrino events.

The West Area Neutrino Facility at CERN provides  a beam of 27 GeV average energy
consisting mainly of $\nu_\mu$ with a  5\% $\bar{\nu}_{\mu}$ contamination. 
During the four  years of operation, 
emulsion target material was 
exposed to the  beam 
with an integrated intensity which corresponds
to $5.06 \times 10^{19}$ protons on target.  The data from the electronic 
detectors were analysed and the events possibly originating from the 
emulsion stacks were identified.

Since the CHORUS experiment was designed primarily to search for  $\nu_\mu\rightarrow
\nu_\tau$ oscillation, the data selection for the first phase of the analysis 
was optimized for the detection of a  $\tau$ decaying into a single charged particle.
It allowed to reach the limit on the $\nu_\mu\rightarrow
\nu_\tau$ oscillation probability as $P_{\mu\tau}\leq$ 3.4 $\times$ 10$^{-4}$ 
at large $\Delta m^{2}$ value \cite{Esk00}. 
CHORUS has also seen neutrino induced diffractive
$D_{s}^{\star}$ production \cite{Ann98} and charm pair production
in charged-current  (CC) interactions \cite{Del00}. 
In addition, with the calorimeter target, 
neutral-current (NC) production of J/$\psi$ was observed \cite{Esk01} and 
preliminary results on differential cross-sections 
and structure functions on lead were obtained \cite{Old00}. 

Recently, the second phase of CHORUS scanning 
and analysis has been started. 
It is based on new reconstruction
software and improved MC packages. The key point is the 
new emulsion scanning method called 'netscan'.
This method, originally developed  for the DONUT experiment ~\cite{Kod01}, 
consists of recording all track segments within a large angular 
acceptance in a volume surrounding the located vertex position. 
It can improve the limit on the $\nu_\mu \rightarrow
\nu_\tau$ oscillation probability by a factor of 3 to reach the design goal.
In the second phase, a search for charm decays in the emulsion is also performed.
Several thousands of such events are expected to be located. 
Based on a small part of the statistics, 
the $D^{0}$ production cross section has already been measured 
with an accuracy better than 10\% \cite{Kay02}.
The following charm physics is of interest: 
\begin{itemize}
\item meson rates, branching ratios for
different channels, extraction of CKM matrix elements $|V_{cd}|$ and
$|V_{cs}|$, strange sea of the nucleon and fragmentation functions; 
\item charmed baryon production and decays; 
\item diffractive production of $D_{s}^{\star}$ to measure
different branching ratios and determine $|V_{cs}|$; 
\item associated charm production; 
\item rare charm decays. 
\end{itemize}
With the calorimeter as a target, it is possible to measure 
$\mathrm{d}^{2}\sigma/\mathrm{d}x\mathrm{d}y$ and structure functions from 
$\nu_{\mu}$ and $\bar{\nu}_{\mu}$ interactions on lead, to study dimuon production
and to search for rare processes. Potentially the
same tasks can be also fulfilled using the first two magnets of the 
iron muon spectrometer as an active target. 
In addition, with a set of lead, iron, marble and 
plastic targets\footnote{At the end of experiment 
the emulsion target was replaced by this set of 4 targets.} 
placed in the neutrino beam
ratios of CC total cross-sections 
are being determined.

\section{CHORUS MC simulation procedure}
The procedure of MC simulation in the CHORUS
experiment 
consists of modelling the neutrino beam,
the generation of physical events, the simulation of the detector response
and the reconstruction.
\subsection{Simulation of the neutrino beam}
Parent mesons for neutrinos, modelled with FLUKA \cite{Fas97}
by NOMAD \cite{Gug01}, are fed into the
CHORUS GEANT 3.21-based \cite{Gea94} neutrino beam simulator
called GBEAM \cite{Sor95}. It performs tracking of these mesons
through the neutrino tunnel and models their decays into neutrinos.
The output of the simulation is the flavour, the creation-vertex 
and the four-momentum of the neutrino. 
\subsection{Generation of physical events}
Deep-inelastic (DIS) events are simulated with the JETTA \cite{Zuc95} package.
(Quasi)-elastic (QE) processes and resonance production 
are modelled with the RESQUE \cite{Ric96} code. 
Information about these general-purpose packages  
can be found in Section 4.
For specific processes, dedicated generators were developed (Section 5).
All event generators use incoming neutrino information
produced by GBEAM. Their output is similar to the LUND LUJETS common block \cite{Sjo94} 
(particle codes, momenta, vertices and history). 
\subsection{Simulation of the detector response}
The package which simulates the detector response, 
called Eficass (Emulsion FIbers CAlorimeter and Spectrometer Simulation), 
contains a very detailed description of the detector geometry
in the GEANT 3.21 framework. The longitudinal vertex positioning 
is performed with a special geantino-based algorithm
in accordance with the material density the neutrino traverses.
Eficass digitizes GEANT hits in each
active volume of the CHORUS detector subsystem. 
Its output, similar to what is produced by the data acquisition system,  
is sent to the reconstruction program.

\section{General-purpose generators}
\subsection{JETTA}
JETTA (JETs in Tau Analysis) \cite{Zuc95} was written  
for the CHORUS kinematical pre-selection of events by embedding into a general 
DIS generator (LEPTO 6.1 \cite{Ing92}) the formalism 
of Ref. \cite{Alb75}
to describe interactions with a massive charged lepton (5 structure functions). 
The code takes care of the polarization effects of the outgoing lepton 
also with respect to its 
decays by using the TAUOLA \cite{Jad93} package. The rest, 
is plain JETSET \cite{Sjo94} code, tuned on the BEBC
data \cite{Gra83,All84} and it deals reasonably well 
with some charm production aspects and
different interactions (NC and CC). 
JETTA generates DIS events ($W^{2}\ge$2 GeV$^{2}$)
initiated by $\nu_{e}$, $\nu_{\mu}$ or $\nu_{\tau}$ with energies in the
region of 3$\div$300 GeV. Structure 
functions GRV94LO \cite{Glu95} are used by default, other choices are possible 
via steering cards. Nucleon composition of the target and Fermi 
motion are also taken into account.
JETTA is able to simulate events of specific classes
like open charm and dimuon production. Threshold behaviour 
of the cross section due to the non-zero charm quark mass ($m_{c}$) is
introduced but without slow rescaling. The Peterson
longitudinal fragmentation function \cite{Pet83} with 
$\varepsilon=$ 0.072 \cite{Vil99} is used 
for charm quark hadronization.       

JETTA reproduces reasonably well neutrino bubble chamber experimental 
data on charged track multiplicities, transverse shower development,
pion energies and fragmentation functions for hadrons.
It was also checked on our CC, charm and dimuon data 
obtained both with emulsion and calorimeter targets.
\subsection{RESQUE} 
RESQUE \cite{Ric96}, an event generator for RESonance and (QUasi)Elastic 
events, has been derived from a preexisting code. Computation
of cross sections, as well as the routines which 
describe resonance decays have been adapted from the Soudan-II
RSQ generator \cite{Bar87}. 
The main changes concern: \\
- significant extension of the energy range for the calculation of cross-sections
and for the generation of events; \\
- the incorporation of nuclear effects, Fermi motion \cite{Bod81} and
Pauli suppression; \\
- the inclusion of $\nu_{\tau}$ interactions, and $\tau$ decays,
taking into account the polarization of the lepton in the 
final state.\footnote{It is assumed
that the generated $\tau$ has always helicity -1 as obtained
from an extrapolation from DIS case \cite{Alb75}.} \\
Three nucleon formfactors are used in the calculation
of QE CC cross section \cite{Lle72}.
The fourth formfactor which is proportional to  
mass squared of the outgoing lepton, is taken into account in the
case of $\nu_{\tau}$ interactions. 
For elastic NC interactions, the cross sections are
taken from Ref.\cite{Hen79}. 

The original routines from RSQ code \cite{Bar87} are used 
to generate the contribution of 16 baryonic resonances
($N^{\star}$ and $\Delta$) with invariant-mass 
$W \le$ 2 GeV. All resonances and their decays are generated
separately. 
The production cross-sections have been computed in 
Ref.\cite{Rei81} using the FKR \cite{Fey71} semirelativistic
quark model. 
 Total cross sections of QE $\nu$n interactions and 
$\Delta^{++}$ production in $\nu$p-
interactions are found to be in an agreement with experimental data
\cite{Bel78,All80,Bel85}.
\subsection{Verification of CHORUS MC chain}
The combined performance of JETTA and RESQUE generators was checked
on CHORUS data. Fig.1 shows the reconstructed
neutrino and antineutrino energy of CC events 
in the calorimeter. The difference between detector-deconvoluted
data and MC does not exceed 10$\div$15\%. It is partially due to 
imperfections in our simulation of neutrino parent meson 
transportation through the decay tunnel. Charged multiplicity
of tracks predicted in the emulsion is presented in Fig.2. 
The agreement between data and MC\footnote{Both data and MC
were passed through the same reconstruction program.} is quite 
satisfactory.
\subsection{MICKEY}
MICKEY \cite{Old00} is used as a fast MC generator
for CC interactions in the calorimeter.
Instead of doing a full GEANT simulation, it applies parameterized
detector resolution and inefficiencies. These parameterizations
have been obtained from test-beam data and from full detector
simulations.
The parton distribution set used in MICKEY is that of GRV94LO
\cite{Glu95}. Violation of the Callan-Gross relation
is modelled using a parameterization of $R(x,Q^{2})$ taken
from a fit to the world data \cite{Whi90}. 
Target mass effects and non-zero $m_{c}$
(slow rescaling) are taken into account.
The structure functions are multiplied by a nuclear correction
as a function of $x$ according to Ref.\cite{Gom94}. 
QED radiative corrections are calculated following the 
scheme of Ref.\cite{Bar86}.
The generator was tuned and validated using
our calorimeter and emulsion data
\cite{Old00,Mel01}.

\section{Specific generators}
Four generators were developed for charm studies: \\
- ASTRA for diffractive $D_{s}$ production \cite{Mel97}; \\
- QEGEN for QE production of charmed baryons \cite{Dic02}; \\
- MCDIMUON for dimuon - involving charm - production \cite{Lem95,Poe99}; \\
- CCBAR for pair production of charmed particles both
in CC and NC \cite{Del00}. \\
In addition, there is a generator for detailed 
simulation of ``white star kinks'', WSK \cite{Sat01}.
\subsection{ASTRA}
The core of ASTRA \cite{Mel97} consists of a set of functions 
$F_{n}$ describing the matrix elements for diffractive production
of (vector) $D_{s}^{\star}$ and direct production of (pseudoscalar) $D_{s}$, 
in neutrino interactions, according to the choice of theoretical models 
indexed by $n$. It is left to the user to select her/his favourite model.
By default, models from Ref.\cite{Gai76} (Ref.\cite{Rei83})
are used for $D_{s}^{\star}$ ($D_{s}$) production. 
The first approach is based on generalized vector dominance 
while the second one uses PCAC as a fundamental ingredient. 
Non-zero $m_{c}$ is taken into account in both. 

It should be noted that a correctly selected integration algorithm 
and correctly computed integration limits are crucial.
Ref.\cite{Che77} contains a thorough study 
of physical and computational pitfalls to be avoided. 

Fragmentation and decays are performed with JETSET \cite{Sjo94}.
Nuclear effects other than an {\it ad-hoc} diffractive slope
are not included in ASTRA.
\subsection{QEGEN}
In the QE charm production a valence
$d$-quark is transformed, through weak interaction, into a
$c$-quark. The other two valence quarks in the nucleon behave
as spectators, so only one hadronic state (charmed baryon
$\Lambda_{c}^{+}$, $\Sigma_{c}^{+}$, $\Sigma_{c}^{++}$ or
$\Sigma_{c}^{\star +}$, $\Sigma_{c}^{\star ++}$) is
produced and finally each event will contain a $\Lambda_{c}^{+}$ baryon.
There are two classes of theoretical models which try to describe
the QE charm production. The first class is based on
$SU(4)$ flavour symmetry which is badly broken. A different 
approach is based on local duality in $\nu$N scattering 
on the basis of QCD. Although the predictions of the models for the
total cross sections differ by one order of magnitude, 
$Q^{2}$-distributions are very similar. 
To simulate the dynamics of QE 
charmed baryon production, the
differential cross sections from Ref.\cite{Shr76} are used.
JETSET \cite{Sjo94} is employed to perform fragmentation
and charmed baryon decays. The model to generate Fermi
motion in emulsion target nuclei is adopted from Ref.\cite{Bod81}.
Note that experimental
data on QE charm production are very poor.
\subsection{MCDIMUON}
The event generator MCDIMUON \cite{Lem95}
was originally developed to simulate opposite-sign dimuon events in 
the CHARM-II detector \cite{Vil99}. CTEQ3L \cite{Lai95} parton distribution
functions are used. The masses of the charm quark and the nucleon
are not neglected in the hard-scattering cross-section
(slow rescaling is included). The fragmentation of
charm quarks is described by the fragmentation function from
Ref.\cite{Pet83}. Excited charmed hadrons are not taken
into consideration. The transverse momentum distribution is chosen
as an exponential function of  $p_{t}^{2}$. A correction for radiative effects
is calculated for the inclusive CC cross-section
using the prescription of Ref.\cite{Bar86}. Decays of
charmed hadrons only depend on the phase space. The
nucleon content of the target is also taken into account.

The code is very flexible and the user can provide many parameters
through steering cards. It was tuned for simulation
of dimuon events in the CHORUS calorimeter \cite{Poe99}. 
MCDIMUON was verified both on CHARM-II and CHORUS 
data.
\subsection{CCBAR}
The simulation of charm pair production in NC and CC interactions is
performed using the general-purpose HERWIG event generator \cite{Mar92}, version 6.1
\cite{Cor99}. 
Several features of this generator, like heavy flavour hadron
production and decay via QCD coherence effects and the cluster
hadronization of the jets via non-perturbative gluon splitting, are
especially relevant. 
The parton-shower approach is used both for initial and final states. 
MRS 5 (Owens) structure functions are used \cite{Mar93}.
The Peterson's model for fragmentation functions is used with 
$\varepsilon=$ 0.072. 
Nuclear effects such as evaporation and re-interactions are not 
simulated.

The process in NC is based on the $Z^{0}$-gluon fusion. 
A dedicated mode to produce this process is foreseen by the generator.  
Associated charm production in CC is based on the 
splitting of $c\bar{c}$ quarks from a gluon which is 
radiated through bremsstrahlung of a light quark. 
No dedicated mode is foreseen for such a process which is
therefore obtained among conventional CC interactions. 
It should be noted however that total cross sections predicted by HERWIG
\cite{Sey95} and calculated in Ref.\cite{Hag80} differ by one order of
magnitude.
The
CCBAR generator cannot be validated as the available experimental
data for associated charm production are very poor
\cite{Del00,Alt00,Ush88}. 
\subsection{WSK}
A hadron-nucleus interaction which produces only one
mimimum ionizing particle and no heavily ionizing tracks or
other signs of nuclear break up is usually called ``white star kink'' (WSK).
Note that, this is the main background for 
oscillation searches in CHORUS in ``0$\mu$'' mode \cite{Esk00}, 
which is very difficult to estimate. 

The simulation \cite{Sat01} is composed of two separate parts. The first is the
generation of hadron interactions in emulsion, the second is the simulation
of emulsion response to the different particles. The hadron-emulsion 
interactions have been simulated using FLUKA \cite{Fas97}.
The emulsion response description is a very delicate point.
It is based on the range in emulsion, for which the parameterization is 
obtained by an old experiment \cite{Bar63}.
The generator has been tested and it reproduces 
the CHORUS data \cite{Esk00} quite well. 

\section{Conclusions}
The CHORUS experiment has two general-purpose 
MC event generators which 
were validated using available experimental data.
Tuning of dedicated generators, developed for
charm physics, is well advanced. Currently, a complete set of
CHORUS MC generators is available. 

\begin{figure}[bhtp]
\begin{center}
\includegraphics[bb=0 0 567 600,width=6cm]{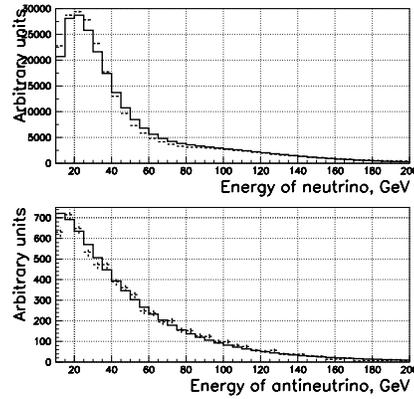}
\vspace*{-2.5cm}
\caption [] {Neutrino (top plot) and antineutrino (bottom plot) 
energy spectra for CC events in the calorimeter.
The solid histograms are for experimental data, the dashed histograms 
for MC.}
\label{Fig1}
\end{center}
\end{figure}

\begin{figure}[bhtp]
\begin{center}
\includegraphics[bb=0 0 567 450,width=5cm]{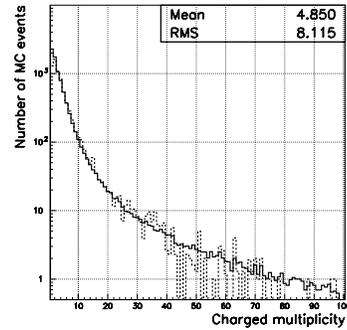}
\vspace*{-2.4cm}
\caption [] {Charged multiplicity for events having
predictions of
particle trajectories into the  emulsion stack for the location of
the vertex positions. The solid histogram is for experimental data, 
the dashed histogram for MC.}
\label{Fig2}
\end{center}
\end{figure}

\end{document}